\documentclass{article}
\usepackage{amsfonts}
\usepackage{amssymb}
\usepackage{amsmath}
\usepackage{amsthm}
\usepackage[english]{babel}
\usepackage{epigraph}

\begin{document}

\title{Weak Gravity Limit in Newer General Relativity}

\author{Alexey Golovnev$^1$, Sofia Klimova$^{2,3}$, A.N. Semenova$^3$, V.P. Vandeev$^3$\\
{\small \it $^1$Centre for Theoretical Physics, The British University in Egypt,}\\
{\small \it El Sherouk City, Cairo 11837, Egypt}\\
{\small  agolovnev@yandex.ru}\\
{\small \it $^2$Peter the Great St. Petersburg Polytechnic University,}\\
{\small \it Saint-Petersburg 195251, Russia}\\
{\small sonia.klim@yandex.ru}\\
{\small \it $^3$Petersburg Nuclear Physics Institute of National Research Centre ``Kurchatov Institute'',}\\ 
{\small \it Gatchina 188300, Russia}\\
{\small ala.semenova@gmail.com \hspace{50 pt} vandeev{\_}vp@pnpi.nrcki.ru} } 
\date{}

\maketitle

\begin{abstract}

We analyse linearised field equations around the Minkowski metric with its standard flat parallel transport in models of Newer GR, that is quadratic actions in terms of nonmetricity tensor. We show that half of the freedom in choosing the model parameters is immediately fixed by asking for reasonable properties of tensors and vectors, defined with respect to spatial rotations, and accurately describe the much more complicated sector of scalars. In particular, we show that, from the teleparallel viewpoint, the model of STEGR with addition of a gradient squared of the metric determinant exhibits three halves new dynamical modes, and not just one like it was previously claimed.

\end{abstract}

\section{Introduction}

Modified teleparallel gravity models are very popular nowadays, both for fundamental research in gravity and for application to cosmology. Dynamical properties of these models are not very clear yet, except for the case of General-Relativity-equivalent options which are nothing but the standard, Einstein-Hilbert GR rewritten in terms of a different action functional. The standard Einstein-Hilbert action is the only one available in the metric geometry for equations of GR when keeping up with the full diffeomorphism invariance at the level of the Lagrangian density. So, one might ask how the new approaches can be claimed to be more fundamental.

The price to pay is to assume that the spacetime manifold is globally parallelisable, or even better, topologically trivial, and introduce yet another parallel transport, a flat one. The required Lagrangian can then be written in terms of torsion \cite{telrev} and/or nonmetricity \cite{1sym, coinsym, trinity} of the flat connection. Note that we anyway need a metric tensor for describing the observational aspects of gravity. The Lagrangian density can then be fine-tuned in such a way that the corresponding field equation is just the same as the Einstein equation for the metric tensor and with no restriction on the flat parallel transport whatsoever. In other words, it is a more involved reformulation of essentially the very same theory \cite{metrinity}.

In the philosophy of trinity \cite{trinity}, people usually like to discuss two particular cases of teleparallel worlds: the metric one, i.e. with torsion but no nonmetricity, and the symmetric one, i.e. with nonmetricity but no torsion. In any case, a flat connection can be represented as a field of tangent space bases (or tetrads, or Vielbeine, or frames...) composed of covariantly constant vectors \cite{geomW}. If the flat connection is metric-compatible, then such tetrad can be chosen everywhere orthonormal, that is a familiar type of choice for most practitioners from other sides of gravity research. At the same time, if the flat connection is symmetric, then the basis can be viewed as a coordinate one for a special, preferred type of coordinates which then play the role of the Cartesian ones \cite{meRM}. In other words, the GR-equivalent models do introduce some additional geometric structures on top of the usual metric for no new predictions \cite{metrinity}.

For sure, it is much more interesting to try modifications of teleparallel equivalents of GR, even though there are many troubles related to this endeavour. Historically, there are two basic ways of doing so. Recent activity has been centered mostly around non-linear modifications of the GR-equivalent Lagrangian densities, in analogy to famous $f(R)$ gravity framework and starting from the $f(T)$ gravity \cite{fT}. Later this approach was also generalised by adding boundary terms to the arguments of a non-linear function \cite{Baha} which actually means nothing but mixing Riemannian and teleparallel objects inside the function. Another, more classical approach \cite{NewGR} which is much older and is getting revived nowadays is about modification of the quadratic fine-tuned action without introducing more non-linearities into it.

It is not just an accident that the models mentioned above, the $f(T)$ gravity \cite{fT} and New GR \cite{NewGR}, are torsion-based. This is the historical version of teleparallel theories dating back to Einstein himself. Symmetric teleparallelism was created much later \cite{1sym,hist2}. Nowadays it is widely used in modified gravity research. We cannot give a full reference list for this very active field here. Let us simply mention, with a few examples, that there are many directions available: unveiling the geometric foundations \cite{meRM}, studying gravitational waves in non-linear extensions of STEGR \cite{Cap1}\footnote{Doing it for a pure $f(Q)$ model \cite{Cap2} feels like a bit of an overshoot though, for the scalar $\mathbb Q$ is in itself quadratic in perturbations around $\eta_{\mu\nu}$ thus automatically reducing the quadratic action to that of the GR-equivalent $\mathbb Q$ for any smooth function $f$.}, discussing the so-called geometric trinity beyond the simplest cases \cite{Cap3}, investigating possible cosmological solutions \cite{QcosmT}, or even doing phenomenology of those \cite{andro}. It all means that we need a better understanding of the foundational aspects of this approach, starting from the simplest features of its weak gravity regime.

We would like to analyse the weak gravity limit of a quadratic symmetric teleparallel theory known as Newer GR \cite{coinsym}, named after an analogous construction in metric teleparallel approach which was called New GR \cite{NewGR}. Since in symmetric teleparallelism one can always say that partial derivatives of the metric tensor components are just their (flat) covariant derivatives written in the preferred coordinates, it is simply a quadratic theory of a symmetric rank-two tensor field in a Minkowski spacetime. We all know that GR is virtually the only available theory of a massless spin two particle which is formally all right for the standard approach to quantum field theories, at the linear level at least. However, in this paper, we are to discuss the Newer GR models from the viewpoint of purely classical equations, and see what can be said about it from the linearised weak gravity regime already.

The plan of our paper is as follows. In Section 2 we give a brief introduction to the models at hand, then turn to the weak gravity limit in Section 3. Impossibility of having a ghost-free model with all metric components dynamical is discussed in Section 4. After that we present the weak gravity equations in Section 5, and the tensor and vector sectors' behaviour is immediately obvious, while the scalars are described in two following sections. Section 6 is devoted to a review of pathological models, while Section 7 counts the degrees of freedom in potentially viable theories, in particular clarifying the situation with a unimodular-like extension of STEGR. In Section 8 we conclude.

\section{The structure of Newer GR}

We assume that the spacetime has a flat symmetric connection, on top of the usual metric and its Levi-Civita connection which is always there \cite{metrinity}. Moreover, we will work in what is called a coincident gauge. Namely, for such a connection, it is always possible to find coordinates in which all connection coefficients vanish \cite{lost-and-found}. In these coordinates, and for this flat connection, all covariant derivatives coincide with the partial ones. In particular, the simple derivative
\begin{equation}
\label{nonmetr}
Q_{\alpha\mu\nu}=\partial_{\alpha}g_{\mu\nu}
\end{equation}
defines the non-metricity tensor $Q_{\alpha\mu\nu}=Q_{\alpha\nu\mu}$. And we also define two vectors
\begin{equation}
\label{nmvec}
Q_{\mu}=Q_{\mu\alpha}^{\hphantom{\mu\alpha}\alpha}, \qquad {\tilde Q}_{\mu}=Q_{\alpha\hphantom{\alpha}\mu}^{\hphantom{\alpha}\alpha}
\end{equation}
constructed from the nonmetricity tensor (\ref{nonmetr}) in the usual way. Operations of raising and lowering the indices are performed by using the metric as always.

The most general (parity-preserving) action of Newer GR \cite{coinsym} can be given as
\begin{equation}
\label{NGRact}
S= \int d^4 x \sqrt{-g}\ {\mathfrak Q}=\frac12\int d^4 x \sqrt{-g} \left(\frac{a_1}{2} Q_{\alpha\mu\nu} Q^{\alpha\mu\nu} - a_2 Q_{\alpha\mu\nu} Q^{\mu\alpha\nu} -\frac{a_3}{2} Q_{\mu}Q^{\mu} +a_4 {\tilde Q}_{\mu}{\tilde Q}^{\mu}+ a_5 Q_{\mu}{\tilde Q}^{\mu} \right)
\end{equation}
in terms of the nonmetricity tensor (\ref{nonmetr}) and vectors (\ref{nmvec}) defined above. The case  of Symmetric Teleparallel Equivalent of General Relativity (STEGR) is 
$$a_1=a_2=a_3=a_5=1,\qquad a_4=0$$
when the nonmetricity scalar $\mathfrak Q$ appears to be the usual $\mathbb Q$ of STEGR. The convenience of discussing this limit is precisely the reason why we have rescaled the coefficients in the action (\ref{NGRact}) compared to many other works.

We can now define the non-metricity conjugate, also often called superpotential, ${\mathfrak P}^{\alpha\mu\nu}$ via 
$${\mathfrak Q}=\frac12 {\mathfrak P}^{\alpha\mu\nu} Q_{\alpha\mu\nu},$$
as a symmetric (${\mathfrak P}^{\alpha\mu\nu}={\mathfrak P}^{\alpha\nu\mu}$) tensor
\begin{equation}
\label{nmconj}
{\mathfrak P}^{\alpha\mu\nu} =  \frac{a_1}{2} Q^{\alpha\mu\nu} - \frac{a_2}{2} \left(Q^{\mu\nu\alpha} + Q^{\nu\mu\alpha}\right) - \frac{a_3}{2} Q^{\alpha} g^{\mu\nu} +\frac{a_4}{2}\left(g^{\alpha\mu}{\tilde Q}^{\nu} + g^{\alpha\nu}{\tilde Q}^{\mu}\right)  +\frac{a_5}{4}\left(2{\tilde Q}^{\alpha} g^{\mu\nu} + g^{\alpha\mu}Q^{\nu} + g^{\alpha\nu}Q^{\mu}\right)
\end{equation}
which defines the quadratic form of the non-metricity scalar.

One can check {\cite{meRM}}, term by term, that it has a nice property of
$$Q^{\alpha\mu\nu}\cdot \delta {\mathfrak P}_{\alpha\mu\nu}= {\mathfrak P}^{\alpha\mu\nu}\cdot \delta Q_{\alpha\mu\nu},$$
and therefore the variation in the action (\ref{NGRact}) works as
$$\delta ( {\mathfrak P}^{\alpha\mu\nu} Q_{\alpha\mu\nu}) = 2 {\mathfrak P}^{\alpha\mu\nu} \cdot \partial_{\alpha} \delta g_{\mu\nu} - \left({\mathfrak P}^{\mu\alpha\beta} Q^{\nu}_{\hphantom{\nu}\alpha\beta} +{\mathfrak P}^{\alpha\mu\beta} Q_{\alpha\hphantom{\nu}\beta}^{\hphantom{\alpha}\nu} + {\mathfrak P}^{\alpha\beta\mu} Q_{\alpha\beta}^{\hphantom{\alpha\beta}\nu}  \right)\cdot \delta g_{\mu\nu}$$
where the last two terms in the brackets are written separately from each other for nothing but clarity of what has been done.

Note also \cite{meRM} that the expression inside the brackets above, ${\mathfrak P}^{\mu\alpha\beta} Q^{\nu}_{\hphantom{\nu}\alpha\beta} +2{\mathfrak P}^{\alpha\beta\mu}Q_{\alpha\beta}^{\hphantom{\alpha\beta}\nu} $, is automatically $\mu\leftrightarrow\nu$ symmetric, so that the equations of motion, in the form of $-\frac{2}{\sqrt{-g}}\cdot \frac{\delta S}{\delta g_{\mu\nu}}$, can safely be written as
\begin{equation}
\label{feq1}
\frac{2}{\sqrt{-g}}\partial_{\alpha}\left(\sqrt{-g} {\mathfrak P}^{\alpha\mu\nu}\right) + {\mathfrak P}^{\mu\alpha\beta} Q^{\nu}_{\hphantom{\nu}\alpha\beta} +2{\mathfrak P}^{\alpha\beta\mu} Q_{\alpha\beta}^{\hphantom{\alpha\beta}\nu}  - {\mathfrak Q} g^{\mu\nu}=0.
\end{equation}
One can also easily transform the upper-indices equation (\ref{feq1}) to the mixed-position form
\begin{equation}
\label{feq2}
\frac{2}{\sqrt{-g}}\partial_{\alpha}\left(\sqrt{-g} {\mathfrak P}^{\alpha\mu}_{\hphantom{\alpha\mu}\nu}\right) + {\mathfrak P}^{\mu\alpha\beta} Q_{\nu\alpha\beta} - {\mathfrak Q} \delta^{\mu}_{\nu}=0
\end{equation}
which is the simplest one and the most convenient for cosmology, or to arguably the most familiar approach
\begin{equation}
\label{feq3}
\frac{2}{\sqrt{-g}}\partial_{\alpha}\left(\sqrt{-g} {\mathfrak P}^{\alpha}_{\hphantom{\alpha}\mu\nu}\right) +{\mathfrak P}_{\mu}^{\hphantom{\mu}\alpha\beta} Q_{\nu\alpha\beta}- 2{\mathfrak P}_{\alpha\beta\nu} Q^{\alpha\beta}_{\hphantom{\alpha\beta}\mu} - {\mathfrak Q} g_{\mu\nu}=0
\end{equation}
with all the indices down.

These are the equations. Then comes the task of analysing physical properties of the theories at hand. Unfortunately, if anything of modified teleparallel models is understood well, it is that they are problematic in many respects \cite{fiss, medegr}. Already the Hamiltonian analysis of $f(T)$ gravity appears to be a complicated topic \cite{torham1, torham2, torham3}, with no proper discussion of strong coupling issues in the Hamiltonian language as of now. A much easier problem of New GR was studied from this point of view only very recently \cite{NGRham, NGRham2}. Many questions still remain open, up to the point that, when a naive count of degrees of freedom in Type 7 New GR\footnote{It is what is called Type 8 in our papers \cite{weNGR1, weNGR2}. It turnes out that we have swapped types 7 and 8 in our New GR notations, compared to other papers \cite{oNGR1, oNGR2}.} results in a negative number of $-1$, it is simply interpreted as no dynamical modes at all \cite{NGRham2} with no more comments whatsoever. Symmetric teleparallel models are much less studied than even that, and related Hamiltonian works go all the way to claiming that the standard algorithm fails \cite{aD}.

As a first step to better understanding a theory of gravity, it is reasonable to analyse its weak gravity limit first, i.e. small perturbations around the trivial background. In models of $f(T)$ and $f(Q)$ gravity, it is not very interesting since all the new stuff is in the strong coupling regime there, so that nothing but the standard GR can be seen. However, the New and Newer GR theories are different in this respect, even though some strong coupling issues are still there \cite{weNGR2}. In our previous papers \cite{weNGR1, weNGR2}, the weak gravity limit of New GR has been studied, and now we turn to the Newer GR case, not restricting ourselves to the principle symbol only \cite{Ulmod}. One might also want to pay attention to works on primary Hamiltonian constraints in Newer GR \cite{Lavinia} and in general quadratic teleparallel \cite{Daniel}, and maybe to very recent investigations in Hamiltonian metric teleparallel \cite{newHam} and linear perturbations in symmetric teleparallel \cite{newSym}.

\section{The weak gravity limit}

In the weak gravity limit, we consider small perturbations around the Minkowski metric in Cartesian coordinates, 
$$g_{\mu\nu}=\eta_{\mu\nu}+h_{\mu\nu},$$ 
and parametrise them in the usual cosmological-perturbations-like manner \cite{cosmopert}:
$$(1+2\phi)\ dt^2 + 2 \left(\partial_i V + V_i \right) dx^i dt - \left((1-2\psi)\delta_{ij} +2 \partial^2_{ij}\sigma +\partial_i c_j +\partial_j c_i + {\mathfrak h}_{ij}\right) dx^i dx^j,$$
or in other words
\begin{equation}
\label{pert}
h_{00}=2\phi, \qquad h_{0i}=V_i+\partial_i V, \qquad h_{ij}=-{\mathfrak h}_{ij}-\partial_i c_j -\partial_j c_i -2 \partial^2_{ij}\sigma+2\psi\delta_{ij},
\end{equation}
with the standard restrictions on the variables, that is $\partial_i V_i=\partial_i c_i=\partial_i {\mathfrak h}_{ij}={\mathfrak h}_{ii}=0$, in order to fully separate, in the linear order, scalars ($\phi$, $V$, $\psi$, $\sigma$), vectors ($V_i$, $c_i$), and tensors (${\mathfrak h}_{ij}$) from each other. A quadratic model is always kind of simple. Below we specify each term in the action to the quadratic limit around $\eta_{\mu\nu}$ and derive their contributions to the linearised equations.

Of course, we could simply use equations in any form (\ref{feq1}, \ref{feq2}, \ref{feq3}). In any case, the linearised weak gravity equation in vacuum can be taken as
\begin{equation}
\label{leq}
\partial_{\alpha} {\mathfrak P}^{\alpha}_{\hphantom{\alpha}\mu\nu}=0,
\end{equation}
where the position of indices doesn't matter in the linear order. However, it is also instructive to derive the shape of equations (\ref{leq}) explicitly. Note that we do not substitute the parametrisations (\ref{pert}) right into the action (\ref{NGRact}), we do it directly into the equations (\ref{leq}), so that everything is safe in this respect, even though substitutions with spatial derivatives only would anyway change the contents of equations only beyond our approach to perturbation theory in which we solve every equation of the form $\bigtriangleup f=0$ simply as $f=0$.

We define the generalisation of the linearised Einstein tensor as
\begin{equation}
\label{genein}
{\mathfrak G}_{\mu\nu}\equiv \frac{1}{\sqrt{-g}} \frac{\delta S}{\delta h^{\mu\nu}}= a_1  {\mathfrak G}^{(1)}_{\mu\nu} + a_2  {\mathfrak G}^{(2)}_{\mu\nu} + a_3  {\mathfrak G}^{(3)}_{\mu\nu} + a_4  {\mathfrak G}^{(4)}_{\mu\nu} + a_5  {\mathfrak G}^{(5)}_{\mu\nu}
\end{equation}
and calculate it as a sum of different terms' contributions (\ref{NGRact}). Note that, with our sign convention, $(+,-,-,-)$, the Einstein-Hilbert Lagrangian density would be equal to $ - R$, and therefore the Einstein tensor would be given by $- \frac{\delta S_{\mathrm{EH}}}{\delta g^{\mu\nu}}$ or equivalently by $ \frac{\delta S_{\mathrm{EH}}}{\delta g_{\mu\nu}}$, while $h_{\mu\nu}=\delta g_{\mu\nu}$.

The {\bf first term} (\ref{NGRact})
$$\frac14 \sqrt{-g}\cdot Q_{\alpha\mu\nu} Q^{\alpha\mu\nu} \approx \frac14 (\partial_{\alpha} h_{\mu\nu}) \partial^{\alpha} h^{\mu\nu}$$
produces
\begin{equation}
\label{eq1}
{\mathfrak G}^{(1)}_{\mu\nu}=-\frac12 \square h_{\mu\nu} + {\mathcal O}(h^2),
\end{equation}
and therefore
\begin{eqnarray*}
{\mathfrak G}^{(1)}_{00} & = & -\square\phi,\\
{\mathfrak G}^{(1)}_{0i} & = & -\frac12 \square \left(V_i + \partial_i V\right),\\
{\mathfrak G}^{(1)}_{ij} & = & \square \left(\frac12  {\mathfrak h}_{ij} + \frac12 \left(\partial_i  c_j + \partial_j  c_i\right) - \psi \delta_{ij} +\partial^2_{ij} \sigma\right)
\end{eqnarray*}
in the linearised limit.

The {\bf second term} (\ref{NGRact})
$$-\frac12 \sqrt{-g}\cdot Q_{\alpha\mu\nu} Q^{\mu\alpha\nu} \approx -\frac12 (\partial_{\alpha} h_{\mu\nu}) \partial^{\mu} h^{\alpha\nu}$$
produces
\begin{equation}
\label{eq2}
{\mathfrak G}^{(2)}_{\mu\nu}=\frac12 \partial^{\alpha}\left(\partial_{\mu}  h_{\alpha\nu} +  \partial_{\nu}  h_{\alpha\mu}\right) + {\mathcal O}(h^2),
\end{equation}
and therefore
\begin{eqnarray*}
{\mathfrak G}^{(2)}_{00}  & = & 2\ddot\phi - \bigtriangleup \dot V,\\
{\mathfrak G}^{(2)}_{0i}  & = & \frac12 \left({\ddot V}_i + \bigtriangleup{\dot c}_i\right) +\partial_i \left(\frac12 \square V + \dot\phi - \dot\psi + \bigtriangleup\dot\sigma \right),\\
{\mathfrak G}^{(2)}_{ij}  & = &  \frac{1}{2} \left(\partial_i \left( {\dot V}_j + \bigtriangleup c_j \right) + \partial_j \left( {\dot V}_i+ \bigtriangleup c_i\right) \right) + \partial^2_{ij} \left(\dot V - 2\psi +2\bigtriangleup\sigma \right)
\end{eqnarray*}
in the linearised limit.

The {\bf third term} (\ref{NGRact})
$$-\frac14 \sqrt{-g}\cdot Q_{\mu} Q^{\mu} \approx -\frac14 (\partial_{\mu} h^{\alpha}_{\alpha}) \partial^{\mu} h^{\beta}_{\beta}$$
produces
\begin{equation}
\label{eq3}
{\mathfrak G}^{(3)}_{\mu\nu}=\frac12 \eta_{\mu\nu} \square h^{\alpha}_{\alpha} + {\mathcal O}(h^2),
\end{equation}
and therefore
\begin{eqnarray*}
{\mathfrak G}^{(3)}_{00}  & = & \square\left(\phi-3\psi+\bigtriangleup\sigma\right),\\
{\mathfrak G}^{(3)}_{0i}  & = & 0,\\
{\mathfrak G}^{(3)}_{ij}  & = & -\delta_{ij}\square\left(\phi-3\psi+\bigtriangleup\sigma\right)
\end{eqnarray*}
in the linearised limit.

The {\bf fifth term} (\ref{NGRact})
$$\frac12 \sqrt{-g}\cdot Q_{\mu} {\tilde Q}^{\mu} \approx \frac12 (\partial_{\mu} h^{\alpha}_{\alpha}) \partial_{\beta} h^{\beta\mu}$$
produces
\begin{equation}
\label{eq5}
{\mathfrak G}^{(5)}_{\mu\nu}=-\frac12\left(\partial_{\mu}\partial_{\nu} h^{\alpha}_{\alpha} + \eta_{\mu\nu} \partial_{\alpha\beta} h^{\alpha\beta}\right) + {\mathcal O}(h^2),
\end{equation}
and therefore
\begin{eqnarray*}
{\mathfrak G}^{(5)}_{00}  & = & -\left(2\ddot\phi - 3\ddot\psi+\square\bigtriangleup\sigma-\bigtriangleup\dot V + \bigtriangleup\psi\right),\\
{\mathfrak G}^{(5)}_{0i}  & = & -\partial_i \left(\dot\phi - 3\dot\psi+ \bigtriangleup\dot\sigma \right),\\
{\mathfrak G}^{(5)}_{ij}  & = & - \partial^2_{ij}\left(\phi - 3\psi+ \bigtriangleup\sigma\right)+\delta_{ij}\left(\ddot\phi -\bigtriangleup\dot V +\bigtriangleup\psi - \bigtriangleup^2 \sigma\right)
\end{eqnarray*}
in the linearised limit.

\subsection{On the term which was neglected above}

Above we have neglected the fourth term (\ref{NGRact}). The reason is that it gives us nothing new in the linearised limit. Indeed, this term
$$\frac12 \sqrt{-g}\cdot {\tilde Q}_{\mu} {\tilde Q}^{\mu} \approx \frac12 (\partial_{\alpha} h^{\alpha\mu}) \partial^{\beta}h_{\beta\mu}$$
can be integrated by parts to
$$\frac12 (\partial_{\alpha} h_{\beta\mu})\partial^{\alpha} h^{\beta\mu}\approx \frac12 \sqrt{-g} Q_{\alpha\beta\mu}Q^{\beta\alpha\mu},$$
and therefore, in the linearised limit, it coincides with minus the second term (\ref{NGRact}).

It is the first sign of the strong coupling issues, in the generalised meaning \cite{mestr} at least. Let us consider a model with 
$$a_1=a_3=a_5=0\qquad \mathrm{and} \qquad a_2 = a_4 \neq 0.$$ 
Non-linearly, its equations of motion are of first order in derivatives only, however they are non-trivial. At the same time, in the linear limit, it is an empty model, with everything being just pure gauge. One can see this at the level of equations (\ref{feq3}) directly. Indeed, given the non-metricity conjugate of the form
$${\mathfrak P}_{\alpha\mu\nu} \propto Q_{\mu\nu\alpha} + Q_{\nu\mu\alpha} - g_{\alpha\mu}{\tilde Q}_{\nu} - g_{\alpha\nu}{\tilde Q}_{\mu} = \partial_{\mu}g_{\alpha\nu} + \partial_{\nu}g_{\alpha\mu} - g_{\alpha\mu} g^{\beta\rho} \partial_{\beta} g_{\rho\nu} - g_{\alpha\nu}g^{\beta\rho} \partial_{\beta} g_{\rho\mu},$$
we get
$$\partial_{\alpha}\left(\sqrt{-g}{\mathfrak P}^{\alpha}_{\hphantom{\alpha}\mu\nu}\right) \propto \partial_{\alpha} \left(\sqrt{-g}g^{\alpha\beta}\left(\partial_{\mu}g_{\beta\nu} + \partial_{\nu}g_{\beta\mu} \right)\right) - \partial_{\mu} \left(\sqrt{-g}g^{\beta\rho} \partial_{\beta} g_{\rho\nu}\right) - \partial_{\nu}\left(\sqrt{-g}g^{\beta\rho} \partial_{\beta} g_{\rho\mu}\right) = {\mathcal O}(h^2)$$
and also see that the second derivatives cancel each other in the full non-linear equations, too.

Note in passing that what was called Type 1 Newer GR,
$$1=a_1=a_3=a_5\neq a_2=a_4$$
in the recent paper \cite{ManVas} modifies the STEGR equations in the lower derivative part only, and what is more worrisome, only by non-linear terms around Minkowski. Therefore, it is prone to strong-coupling issues. It would not be the case if the whole dynamical structure was the same in the two. However, the STEGR action is the only one with the full diffeomorphism invariance in it.

All in all, the linearised equations (\ref{genein}) take the reduced form
\begin{equation}
\label{redein}
{\mathfrak G}_{\mu\nu}= a_1  {\mathfrak G}^{(1)}_{\mu\nu} + (a_2 - a_4)  {\mathfrak G}^{(2)}_{\mu\nu} + a_3  {\mathfrak G}^{(3)}_{\mu\nu} + a_5  {\mathfrak G}^{(5)}_{\mu\nu},
\end{equation}
and we define the new coefficient
\begin{equation}
\label{ncof}
{\tilde a}_2\equiv a_2 - a_4
\end{equation}
to be used in all the formulae below.

\section{Impossibility of a fully dynamical ghost-free model}

Given the flat connection in the foundations of the approach, it would be reasonable to look for a model with no gauge freedom of choosing it. It requires a serious modification of GR-equivalent models, and there are many doubts of whether it is feasible to get one, in a stable and physical way. The constraint structure of a highly non-linear model can often exhibit bifurcations \cite{Nest} leading to the number of degrees of freedom being ill-defined, often around physically interesting backgrounds.

One possible way to avoid such troubles is to ensure that the kinetic matrix is everywhere non-degenerate and therefore all the variables are truly dynamical. Both New GR and Newer GR are the most general quadratic in velocities actions inside their classes, and therefore making the matrix non-degenerate isn't difficult. However, another question is whether the result would be any stable. Already in New GR, there are various opinions about that \cite{medegr, difstab}. We can take \cite{medegr} the metric teleparallel models as theories of four vector fields with kinetic parts in terms of ${\mathcal F}^a_{\mu\nu}=\partial_{\mu}e^a_{\nu}-\partial_{\nu}e^a_{\mu}$, therefore one might hope to get a healthy model with all components, modulo diffeomorphisms, dynamical.

The situation in Newer GR is different. First of all, already in GR itself, the kinetic matrix is not positive definite, and the would-be ghost is killed by a constraint which makes it non-dynamical. Therefore, making a stable fully-dynamical model necessarily requires modifying the action far away from the STEGR case. What we would like to show is that even this does not allow for a ghost-free fully-dynamical version of Newer GR. Of course, it should not be very surprising when we are trying to give dynamics to all tensor field components in a Lorentzian setting.

One can immediately calculate the quadratic in velocities part of the quadratic action as
$$\frac{a_1}{4} \dot h_{\mu\nu} \dot h^{\mu\nu} - \frac{{\tilde a}_2}{2} \dot h_{0\mu} \dot h^{0\mu} -\frac{a_3}{4} (h^{\mu}_{\mu})^2 + \frac{a_5}{2} \dot h^{\mu}_{\mu} \dot h_{00}.$$
By using integrations by parts and Fourier representation of $- V \bigtriangleup V=  k^2 V^2$ for both $V$ and $c_i$, and also treating some collections of variables as one, for a vector of variables
$$\left(\left\{\frac{\dot{\mathfrak h}_{ij}}{2},\ \frac{k \dot c_i}{\sqrt{2}}\right\} \quad \left\{\frac{\dot V_i}{\sqrt{2}},\ \frac{k \dot V}{\sqrt{2}}\right\} \quad \dot\phi \quad \bigtriangleup\dot\sigma \quad \dot\psi\right)$$
one finds the matrix
\begin{equation}
\label{kinmat}
{\mathfrak K}=
\left(\begin{array}{ccccc}
a_1 & 0  & 0  & 0  & 0  \\
0  & {\tilde a}_2 - a_1  & 0  & 0  & 0 \\
0  & 0  & a_1 - 2{\tilde a}_2 -a_3 +2a_5  & -a_3 + a_5  & 3a_3 - 3a_5 \\
0  & 0  & -a_3 + a_5  & a_1 - a_3  & -a_1 + 3a_3 \\
0  & 0  & 3a_3 - 3a_5  &  -a_1 + 3a_3  & 3a_1 - 9a_3 
\end{array}\right).
\end{equation}

Of course it is easy to have it non-degenerate (even with ${\tilde a}_2=a_3=a_5=0$). However, the issue is then about its positive definiteness. The necessary condition of 
$${\tilde a}_2>a_1>0$$ 
is obvious, and then one has to study the most difficult case of scalars (except $V$). We need to make the non-trivial lower right corner of the matrix positive definite. Again, if it was not for the mixed spatiotemporal metric components ($V$ and $V_i$), a possible case would be ${\tilde a}_2=a_3=a_5=0$ and $a_1>0$. We aim at proving that making the whole matrix (\ref{kinmat}) positive definite is not possible.

The Sylvester's criterion tells us that our matrix is positive definite if and only if its leading principle minors (i.e. left upper corners) have positive determinants. Since renumbering the elements does not change the quadratic form, every principle minor (i.e. complementary to any subset of diagonal elements) must have its determinant positive, too. In case of any doubt, we refer the reader to any textbook on matrix algebra. However, for proving the claimed impossibility we only need the positive determinants as a necessary condition, and this is obvious. 

Indeed, let's take a vector in a subspace of $\dot\psi$, for example, then the quadratic form is positive if the corresponding (diagonal) matrix element is positive. Therefore, we need 
$$a_1>3a_3,$$ 
thus already going far away from GR. If we take an arbitrary vector in the subspace of $\bigtriangleup\dot\sigma$ and $\dot\psi$, then the quadratic form is governed by the lower right corner of the matrix, and the corresponding determinant must also be positive, as a product of positive eigenvalues. Together with $a_1>3a_3$, it reproduces $a_1>0$ again.

Let's finally look at what the $\dot\phi$ and $\bigtriangleup\dot\sigma$ subspace needs. With a little bit of simple algebra we find the corresponding determinant
$$(a_1 - 2{\tilde a}_2 -a_3 +2a_5 ) \cdot ( a_1 - a_3 ) - (-a_3 + a_5)^2 = - (a_1 - a_5)^2 - 2 ({\tilde a}_2 -a_1)\cdot (a_1 - a_3) <0$$
negative, due to the previous requirements. Note that it is precisely the positivity in the mixed spatiotemporal components what makes positivity in all the rest impossible.

Summarising it all, a fully-dynamical ghost-free, let alone fully stable, model is not possible. At the same time, neglecting the questions of stability, one can easily find the condition for it being fully-dynamical indeed. By adding three times the penultimate row to the last row of the kinetic matrix (\ref{kinmat}), one gets a matrix whose determinant is very easy to compute:
\begin{equation}
\label{kindet}
\mathrm{det}{\mathfrak K}=2 a^5_1 \left({\tilde a}_2-a_1\right)^3 \left(  a_1^2 - 2a_1 ({\tilde a}_2+ 2a_3-a_5) + 6 {\tilde a}_2 a_3 - 3a_5^2 \right)
\end{equation}
where we have taken into account that the first and the second entries of the matrix represent four and three variables respectively, while the fifth power of $a_1$ comes from the non-trivial part in the left lower corner. As long as the quantity (\ref{kindet}) is not equal to zero, all the modes are dynamical.

\section{Field equations in the weak gravity limit}

In order to analyse the models of Newer GR and classify the possible numbers of degrees of freedom in them, we combine the contributions (\ref{eq1},\ref{eq2},\ref{eq3},\ref{eq5}) in the gravity tensor (\ref{redein}). Let's start from the {\bf tensor sector}. Its only contribution to the linearised $\mathfrak G$ tensor (\ref{redein}) reads
$${\mathfrak G}_{ij} =\frac{a_1}{2}  \square  {\mathfrak h}_{ij}.$$
Therefore, with the equation of
\begin{equation}
\label{teneq}
a_1 \square  {\mathfrak h}_{ij}=0,
\end{equation}
the ${\mathfrak h}_{ij}$ field is never constrained. It is dynamical as long as $a_1 \neq 0,$ and is a pure gauge otherwise.

In the {\bf vector sector} one gets (\ref{redein})
\begin{equation*}
{\mathfrak G}_{0i} =  \frac{{\tilde a}_2 -a_1}{2} \ddot V_i +\frac{a_1}{2}\bigtriangleup V_i +\frac{{\tilde a}_2}{2} \bigtriangleup{\dot c}_i ,
\end{equation*}
\begin{equation*}
{\mathfrak G}_{ij}  =  \frac{a_1}{2}\left( \partial_i \ddot c_j + \partial_j \ddot  c_i\right) + \frac{{\tilde a}_2 - a_1}{2} \bigtriangleup\left(\partial_i  c_j  + \partial_j  c_i \right) + \frac{{\tilde a}_2}{2} \left(\partial_i {\dot V}_j + \partial_j  {\dot V}_i \right)
\end{equation*}
where ${\tilde a}_2 = a_2 - a_4$ as we have introduced above (\ref{ncof}). We immediately see that, even inside the vector sector only, the fully dynamical ghost freedom, 
$${\tilde a}_2 > a_1 >0,$$
leads to gradient instability in both vector fields. Neglecting the stability issues, we can say again that the necessary condition for all vector modes being dynamical is
$$a_1 \neq 0, \qquad {\tilde a}_2\neq a_1.$$

In the perturbative approach, we can write the equation of motion in vacuum as
\begin{eqnarray}
\label{veq1}
({\tilde a}_2-a_1) \ddot V_i + \bigtriangleup \left(a_1 V_i + {\tilde a}_2 \dot c_i\right) & = & 0,\\
\label{veq2}
a_1 \ddot c_i + ({\tilde a}_2 -a_1) \bigtriangleup c_i + {\tilde a}_2 \dot V_i & = & 0.
\end{eqnarray}
The three special cases are very easy to see. In the case of GR, $a_1\neq 0$ and ${\tilde a}_2=a_1$, equations (\ref{veq1}, \ref{veq2}) reduce to $V_i+\dot c_i=0$ which means one constrained mode and one gauge freedom. In non-GR case of $a_1 = 0$ and ${\tilde a}_2\neq 0$, one gets $\dot V_i + \bigtriangleup c_i =0$, with the same nature of modes, differently distributed though. Finally, the vector sector is fully empty, i.e. pure gauge, if $a_1={\tilde a}_2=0$. As we see, unlike in the case of New GR, the vector sector of Newer GR is extremely simple. 

At the same time, we will find below that the {\bf scalar sector} is somewhat more complicated. The ${\mathfrak G}_{\mu\nu}$ tensor components (\ref{redein}) can be presented as 
$${\mathfrak G}_{00}=-(a_1-2{\tilde a}_2-a_3+2a_5)\ddot\phi-3 (a_3-a_5)\ddot\psi + (a_3-a_5)\square\bigtriangleup\sigma - ({\tilde a}_2-a_5)\bigtriangleup\dot V +(a_1-a_3)\bigtriangleup\phi + (3a_3-a_5)\bigtriangleup\psi,$$
$${\mathfrak G}_{0i}=\partial_i\left(\frac{{\tilde a}_2 -a_1}{2} \square V + ({\tilde a}_2 -a_5) \left(\dot\phi + \bigtriangleup\dot\sigma\right) -( {\tilde a}_2 - 3a_5) \dot\psi   \right),$$
\begin{multline*}
{\mathfrak G}_{ij} = \partial^2_{ij} \left(a_1 \ddot\sigma +{\tilde a}_2 \dot V - (a_1- 2{\tilde a}_2 +a_5)\bigtriangleup\sigma -a_5 \phi - (2{\tilde a}_2 
-3a_5)\psi \right) \\
-\delta_{ij}\left((a_1 - 3a_3)\ddot\psi + (a_3- a_5)\ddot\phi + a_3 \bigtriangleup\ddot \sigma + a_5 \bigtriangleup\dot V - (a_1-3a_3+a_5)\bigtriangleup\psi - a_3 \bigtriangleup\phi - (a_3-a_5)\bigtriangleup^2\sigma\right).
\end{multline*}
As always in such perturbation theory, we treat $\bigtriangleup$ as a nonzero number, and also we solve the $\delta_{ik}$ and $\partial^2_{ij}$ parts of the spatial equation separately. Therefore, we put the vacuum equations into the following form:
\begin{eqnarray}
\label{seq1}
({\tilde a}_2 -a_1) \square V + 2({\tilde a}_2 -a_5) \left(\dot\phi + \bigtriangleup\dot\sigma\right) - 2( {\tilde a}_2 - 3a_5) \dot\psi &  = & 0,\\
\label{seq2}
a_1 \ddot\sigma +{\tilde a}_2 \dot V - (a_1- 2{\tilde a}_2 +a_5)\bigtriangleup\sigma -a_5 \phi - (2{\tilde a}_2  - 3a_5)\psi & = & 0,\\
\label{seq3}
(a_1-2{\tilde a}_2-a_3+2a_5)\ddot\phi +  (a_3-a_5) \left(3\ddot\psi - \square\bigtriangleup\sigma\right) +  \bigtriangleup\left(({\tilde a}_2-a_5)\dot V - (a_1-a_3)\phi - (3a_3-a_5)\psi\right) & =& 0,\\
\label{seq4}
(a_3- a_5)\ddot\phi  + (a_1 - 3a_3)\ddot\psi+ a_3 \bigtriangleup \ddot \sigma +\bigtriangleup\left( a_5 \dot V - a_3 \phi  - (a_1-3a_3+a_5)\psi - (a_3-a_5)\bigtriangleup\sigma\right) & = & 0
\end{eqnarray}
where the order is the following: mixed spatiotemporal components, the nondiagonal part of spatial components, the temporal component, the diagonal part of spatial components.

Obviously, the scalars are the challenging part of analysis. Note though that the condition of ${\tilde a}_2=a_1$, which is necessary for any hint at stability in non-scalars, makes the variable $V$ represent one half of a dynamical degree of freedom, at most.

\section{Review of pathological models}

We first briefly review the situation with pathological options. On one hand, as we have seen above in non-scalars, the most general models are necessarily plagued by either ghosts or gradient instabilities already at the level of linearised weak gravity equations. This is the pathological case of $a_1\neq 0$ and ${\tilde a}_2\neq a_1$. On the other hand, there is an option of $a_1=0$ for which the usual two polarisations of gravitational waves are pure gauge, in the linear weak gravity regime at least. Therefore, such models are not interesting as models of gravity. And anyway, it would probably be a bit strange, to use a symmetric tensor field for something which doesn't need the rich structure. Nevertheless, for completeness, it's worth looking at these models' features, too.

\subsection{Unhealthy fully dynamical non-scalar sector}

For all the modes to be fully dynamical, the equations must be uniquely solvable for accelerations. Then, assuming that the previous inequalities, $a_1 \neq 0$ and ${\tilde a}_2\neq a_1$, are satisfied, the first two scalar equations (\ref{seq1}, \ref{seq2}) can immediately be solved for $\ddot V$ and $\ddot\sigma$. This is all we need for the former since this acceleration does not appear anywhere else, while the latter result can be substituted into the last two equations (\ref{seq3}, \ref{seq4}) which then serve as equations for $\ddot\phi$ and $\ddot\psi$. Their kinetic part
\begin{equation}
\label{nmat}
\left(\begin{array}{cc}a_1-2{\tilde a}_2 - a_3 +2a_5 & 3a_3 - 3a_5\\ a_3 - a_5 & a_1 - 3a_3\end{array}\right) \left(\begin{array}{c}\ddot\phi\\ \ddot\psi\end{array}\right) 
\end{equation}
appears then to feature the matrix of determinant $a_1^2 - 2a_1 ({\tilde a}_2+ 2a_3-a_5) + 6 {\tilde a}_2 a_3 - 3a_5^2$ which is precisely the non-trivial minor, modulo the sign reversal of the second row, you get in calculating the determinant (\ref{kindet}) of $\mathfrak K$ by first adding three times its penultimate row to its last row.

Therefore, the condition of equations' (\ref{redein}) solvability for all accelerations is as follows:
\begin{equation}
\label{nonodeg}
a_1 \neq 0, \qquad {\tilde a}_2\neq a_1, \qquad a_5 \neq \frac13 \left(a_1\pm\sqrt{4 a_1^2 -6 a_1 {\tilde a}_2 -  12 a_1 a_3 + 18{\tilde a}_2 a_3}\right).
\end{equation}
In this case, and only in this case, all ten metric components are fully dynamical. Not surprisingly, the scalar sector is the least nice of all, and it is a bit more challenging now than it was in the metric teleparallel \cite{weNGR1}.

In order to review the structure of other possible models, we continue with the case of 
$$a_1\neq 0\quad \mathrm{and}\quad {\tilde a}_2 \neq a_1.$$
It means both tensors and vector are fully dynamical, even though severely unstable. Going away from the general case (\ref{nonodeg}), we make the matrix (\ref{nmat}) in front of $({\ddot\phi}, {\ddot\psi})$ degenerate. 

As the simplest option, the matrix (\ref{nmat}) fully vanishes in case of
$$\frac13 a_1=\frac12 {\tilde a}_2=  a_3=a_5$$
only. In perturbation theory (neglecting the freedom of harmonic functions), the equations (\ref{seq1}, \ref{seq2}, \ref{seq3}, \ref{seq4})
$$\frac12\square V=\dot\phi+\dot\psi+\bigtriangleup\dot\sigma, \quad 3\ddot\sigma +2\dot V=\phi+\psi, \quad \frac12 \bigtriangleup\dot V=\bigtriangleup(\phi+\psi), \quad  \bigtriangleup(\ddot\sigma +\dot V)= \bigtriangleup(\phi+\psi)$$
 reduce to just two independent ones. One can call the variables $\sigma$ and $\phi-\psi$ pure gauge, with two other constrained variables: 
$$V=-2\dot\sigma\qquad \mathrm{and} \qquad \phi+\psi=-\ddot\sigma.$$ 
Note that, barring an empty model with identically vanishing Lagrangian, this case of zero matrix (\ref{nmat}) requires fully dynamical (and unhealthy) tensors and vectors.

What still remains to be uncovered is given by other options of  
\begin{equation}
\label{quadcond}
a_5 = \frac13 \left(a_1\pm\sqrt{4 a_1^2 -6 a_1 {\tilde a}_2 -  12 a_1 a_3 + 18{\tilde a}_2 a_3}\right).
\end{equation}
One of the simplest examples of this sort can be got by generalising the model with the vanishing matrix (\ref{nmat}) by relaxing the condition on ${\tilde a}_2$ only.  Therefore, we assume
$$\frac13 a_1=  a_3=a_5\neq 0$$ 
with ${\tilde a}_2$ being free. By subtracting three times equation (\ref{seq4}) from a Laplacian of equation (\ref{seq2}), we can find a constraint
$$2a_5 \phi - 2({\tilde a}_2-3a_5)\psi + 2({\tilde a}_2-2a_5)\bigtriangleup\sigma + ({\tilde a}_2 -3a_5) \dot V=0,$$ 
Given that there is no $\ddot\psi$ acceleration in any of the equations (\ref{seq1}, \ref{seq2}, \ref{seq3}, \ref{seq4}), it is very tempting to simply exclude the variable $\psi$. However, substituting $\psi=\frac12 \dot V + ...$ into the equation (\ref{seq1}) we kill the acceleration $\ddot V$, so that it doesn't work in a simple way. Instead of that, one can solve for $\dot V$, and after that the equation (\ref{seq1}) gives us yet another constraint:
$$({\tilde a_2}-3a_5)\bigtriangleup V=2({\tilde a}_2-2a_5)\dot\psi + 2a_5\bigtriangleup\dot\sigma.$$
It is then easy to see that its time derivative is a linear combinations of equations (\ref{seq3}) and (\ref{seq4}). Therefore, the system of equations can be reduced to the two constraints above and equation (\ref{seq4}):
$$\ddot\sigma + \dot V = \phi +\psi.$$
In other words, the variable $\sigma$ can be taken as representing gauge freedom, while all the rest then needs two Cauchy data, thus effectively producing one dynamical and two constrained modes.

This example above is very typical of what we see in these models. All in all, it seems that the primary constraint related to the condition (\ref{quadcond}) generically comes accompanied by a gauge freedom. Moreover, it seems to be a very general property of the models with at least one primary constraint. For the other primary constraints, we have seen it in non-scalars and will see its repercussions in the scalars below. This is probably because we have Lorentz-covariant equations in second derivatives of a symmetric tensor only, and it does not allow us to remove the time derivatives without touching upon the spatial ones. Let us not go into details of general analysis since it would be quite cumbersome, while the models at hand are not much interesting physically for always having either ghosts or gradient instabilities already at the level of linear weak gravity equations. 

We finish this subsection by giving an example of analysing a very simple model. It goes with
$${\tilde a}_2=a_3=0\quad \mathrm{and}\quad a_1=a_5\neq 0.$$
The equations take the form of
$$\frac12 \square V = 3\dot\psi -\dot\phi-\bigtriangleup\dot\sigma, \qquad \ddot\sigma - 2\bigtriangleup\sigma=\phi-3\psi \qquad \ddot\phi-\ddot\psi=\frac12 \bigtriangleup\dot V = 2\bigtriangleup\psi - \bigtriangleup^2\sigma.$$
Introducing new variables $\left(X=\phi-\psi,\ Y=2\psi-\bigtriangleup\sigma,\ \sigma,\ V \right)$ and noticing that, given the condition $\frac12 \dot V =Y$, we get the equation $\frac12 \square V=\dot Y - \dot X$ equivalent to $\frac12 \bigtriangleup V = \dot X$ from which $\frac12 \bigtriangleup\dot V = \ddot X$ follows. It all can be rewritten as
$$\frac12 \dot V =Y, \qquad \frac12 \bigtriangleup V = \dot X, \qquad \square\sigma = X-Y.$$
We see that $\sigma$ can be treated as pure gauge, and then the remaining system requires two intial data, therefore behaving like one pure gauge, one dynamical mode, and two constrained modes.

\subsection{No physical tensors}

Another unphysical option is to demand absence of tensor modes at all,
$$a_1=0,$$
and for now we assume ${\tilde a}_2 \neq 0$. Due to acceleration structure of the equations (\ref{seq3}, \ref{seq4}), we introduce a new variable
$$\Sigma\equiv 3\psi - \bigtriangleup\sigma$$
and rewrite the equations (\ref{seq1}, \ref{seq2}, \ref{seq3}, \ref{seq4}) in a simpler form of
\begin{equation}
\label{path1}
{\tilde a}_2  \square V + 2({\tilde a}_2 -a_5) \left(\dot\phi - \dot\Sigma\right)+ 4{\tilde a}_2\dot \psi   =  0,
\end{equation}
\begin{equation}
\label{path2}
{\tilde a}_2 \dot V -a_5 \phi  - (2{\tilde a}_2 -a_5)\Sigma +4 {\tilde a}_2 \psi   =  0,
\end{equation}
\begin{equation}
\label{path3}
(2{\tilde a}_2+a_3-2a_5)\ddot\phi -  (a_3-a_5) \ddot\Sigma -  ({\tilde a}_2-a_5)\bigtriangleup\dot V 
- \bigtriangleup\left(\vphantom{\dot V} a_3\phi - (a_3-a_5)\Sigma - 2a_5\psi\right) =  0,
\end{equation}
\begin{equation}
\label{path4}
(a_3- a_5)\ddot\phi  - a_3\ddot\Sigma + a_5 \bigtriangleup\dot V - \bigtriangleup\left(\vphantom{\dot V} a_3\phi - (a_3-a_5)\Sigma - 2a_5\psi\right)=0.
\end{equation}

The second equation (\ref{path2}) is a constraint. By substituting it into the equation (\ref{path1}), we can find yet another constaint:
\begin{equation}
\label{path1b}
{\tilde a}_2 \bigtriangleup V =(2{\tilde a}_2 -a_5)\dot\phi + a_5 \dot\Sigma.
\end{equation} 
Note that, as long as ${\tilde a}_2 \neq 0$, we could have tried to go another way around and solve the equation (\ref{path2}) for $\psi$. However, even though it does not spoil the dynamical character of the equations (\ref{path3}, \ref{path4}) for $\phi$ and $\Sigma$, substituting it into equation (\ref{path1}) again kills the acceleration $\ddot V$ and produces the constraint (\ref{path1b}). At the same time, the difference of the remaining two equations (\ref{path3}, \ref{path4}) is nothing but the time derivative of the constraint (\ref{path1b}). Therefore, we've got a gauge freedom once more.

At the next step, we look at the system of equations (\ref{path2}, \ref{path4}, \ref{path1b}). If we solve the constraints (\ref{path2}, \ref{path1b}) for $\psi$ and $\bigtriangleup V$ respectively, then the equation (\ref{path4}) transforms into
$$(2{\tilde a}_2 a_3 - a_5^2) \square (\phi-\Sigma)=0.$$
In general, we assume
$$2{\tilde a}_2 a_3 \neq a^2_5,$$
i.e. that the $a_1=0$ incarnation of the condition (\ref{quadcond}) must be violated. Then, $\phi-\Sigma=\phi-3\psi+\bigtriangleup\sigma$ is a dynamical variable, while $\phi+\Sigma$ can be taken as pure gauge, with $V$ and $\psi$ being constrained. Otherwise, if $2{\tilde a}_2 a_3 = a^2_5$, then the equations  (\ref{path2}, \ref{path1b}) is all dynamical information we have, or two pure gauges and two constrained modes only.

Note that we seriously used the condition of ${\tilde a}_2\neq 0$ above. Let's take a look at the case of non-scalars fully removed to gauge freedom,
$$a_1={\tilde a}_2=0.$$
The first two equations (\ref{path1}, \ref{path2}) take the form
$$a_5 \left(\phi-\Sigma\right)=0.$$

If $a_5\neq 0$, then we get from all equations (\ref{path1}, \ref{path2}, \ref{path3}, \ref{path4})
$$3\psi-\bigtriangleup\sigma=\phi \qquad \mathrm{and}\qquad \ddot\phi -\bigtriangleup\dot V - \bigtriangleup(2\psi-\phi)=0$$
for any value of $a_3$. Therefore one can say that $\phi$ and $V$ are pure gauge while $\psi$ and $\sigma$ are fully constrained.

Finally, if $a_1={\tilde a}_2=a_5=0$, so that only $a_3\neq 0$, the only equation we have is
$$\square  \left(\phi-3\psi+\bigtriangleup\sigma\right)=0.$$
It is just one canonical scalar field, with all the rest being pure gauge.

\section{Potentially interesting models}

We turn to the case of 
$$a_1\neq 0\quad  \mathrm{and} \quad {\tilde a}_2=a_1,$$
like in GR. The tensors are dynamical, while the vectors are half gauge and half constrained. Up to now, this is the same as in GR. Only the scalars can still be different.

Since $\square V$ has disappeared from the equation (\ref{seq1}), we immediately get a constraint of
\begin{equation}
\label{constr}
(a_1 -a_5) \left(\dot\phi + \bigtriangleup\dot\sigma\right) = (a_1 - 3a_5) \dot\psi.
\end{equation}
Then the equation (\ref{seq3}) for $\ddot\phi$ and $\ddot\psi$ can be transformed by
$$(-a_1+ a_5-a_3+a_5)\ddot\phi + 3 (a_3-a_5)\ddot\psi =-(a_3-a_5)\ddot\phi -(a_1-3a_3) \ddot\psi + (a_1 - a_5)\bigtriangleup\ddot\sigma,$$
thus reproducing the combination of these accelerations in another equation (\ref{seq4}). In other words, the matrix in front of accelerations becomes degenerate, and at most one combination of $\phi$ and $\psi$ appears to be dynamical, while the difference of the two equations (\ref{seq3}, \ref{seq4}) tells us that 
$$\ddot\sigma+ \dot V = \phi + \psi$$
where we have relied on the assumption that $a_1\neq 0$. Then by substituting this $\ddot\sigma+ \dot V$ into another equation (\ref{seq2}) we get a constraint of
$$(a_1-a_5) \left(\phi+\bigtriangleup\sigma\right)=(a_1-3a_5)\psi$$
which makes the previously found constraint (\ref{constr}) obsolete.

Altogether, we get the system of equations
\begin{eqnarray}
\label{hseq1}
(a_1-a_5) \left(\phi+\bigtriangleup\sigma\right) - (a_1-3a_5)\psi & = & 0,\\
\label{hseq2}
\ddot\sigma+ \dot V -\phi -\psi & = & 0,\\
\label{hseq3}
(a_3- a_5)\ddot\phi  + (a_1 - 3a_3)\ddot\psi+ a_3 \bigtriangleup\ddot \sigma + \bigtriangleup\left( a_5 \dot V - a_3 \phi  - (a_1-3a_3+a_5)\psi - (a_3-a_5)\bigtriangleup\sigma\right) & = & 0
\end{eqnarray}
and can take $\sigma$ as pure gauge, and then three initial data are needed for solving the system, one datum for $V$ and two data for $(a_3- a_5)\phi  + (a_1 - 3a_3)\psi$, so that it behaves as three halves constrained and three halves dynamical modes. This is true as long as the potentially dynamical combination of $\phi$ and $\psi$ is linearly independent of the combination $(a_1-a_5) \phi - (a_1-3a_5)\psi$ constrained by the first equation (\ref{hseq1}), i.e.
\begin{equation}
\label{heldeg}
a_1^2 - 2a_1 a_3 - 2 a_1 a_5 + 3a_5^2 \neq 0
\end{equation}
which is nothing but negation of the condition (\ref{quadcond}) when ${\tilde a}_2=a_1$.

If this inequality is not satisfied, then both $\phi$ and $\psi$ are constrained, with only one half degree of freedom remaining in $V$. An exception to that happens if the whole would-be dynamical equation just disappears. In other words, it might be that, on the constraint surface (\ref{hseq1}), the other two equations (\ref{hseq2}, \ref{hseq3}) just totally coincide with each other, up to an overall factor. A straightforward, though a bit boring, game with quadratic equations shows that it happens if $a_1=a_3=a_5$ only. 

This is a very well-known result. In case of 
$$a_1={\tilde a}_2=a_3=a_5\neq 0$$
we get the linearised GR,
$$\psi=0, \qquad \phi = \ddot\sigma + \dot V, $$
of two pure gauge and two constrained modes in the scalar sector.

\subsection{On the case of a unimodular-like deformation of STEGR}

The results of this section may come as a surprise. The case of breaking only the volume part of diffeomorphism invariance,
$$a_1={\tilde a}_2=a_5\neq 0, \qquad a_3\neq a_1,$$
has been considered \cite{unisym} before\footnote{Strangely enough, in the paper \cite{unisym}, after having introduced the symmetric teleparallel framework in the usual notiations, they suddenly switch to putting an $a_5$ coefficient in from of the 3rd term, apparently borrowing this notation from the previous paper \cite{unisymold}.}, claiming only one new dynamical degree of freedom (compared to GR) and one global free parameter (like the cosmological constant in unimodular gravity). Namely, the model was shown to be equivalent to STEGR with a canonical scalar field of an exponential potential term with an arbitrary constant in front of it \cite{unisym}.

In the perturbative analysis, we are not expected to see global freedoms, however the local dynamical modes must be clear, while what we see is one half dynamical degree of freedom more than what was claimed before. 
More precisely, the equations  (\ref{hseq1}, \ref{hseq2}, \ref{hseq3}) can be reduced to
\begin{equation}
\label{unimsym}
\psi=0, \quad \ddot\sigma + \dot V -\phi =0, \quad \ddot\phi - \bigtriangleup\dot V - \bigtriangleup^2\sigma=0.
\end{equation}
One gauge freedom should be taken in terms of either $\phi$ or $\sigma$, and the remaining system requires then three initial data indeed.

What has happened? It is simply that the previous claim \cite{unisym} is wrong. Not in the scalar-tensor representation, but in how the degrees of freedom count was done. Below we first confirm our result once more. Then we show that their analysis is also correct. Finally, we explain where the extra half degree of freedom comes from.

To check our derivations, one might notice that the models of
$$a_1={\tilde a}_2=a_5, \qquad a_3=a_1\cdot (1+\epsilon)$$
gets the linearised gravity tensor (\ref{redein})
$${\mathfrak G}_{\mu\nu} = G_{\mu\nu} + \epsilon\eta_{\mu\nu} \square (\phi-3\psi+\bigtriangleup\sigma)$$
where $G_{\mu\nu}$ is the usual Einstein tensor, and we have renormalised it all by the common value of the coefficients in the STEGR part. Therefore we get the equations
$$2\bigtriangleup\psi + \epsilon  \square (\phi-3\psi+\bigtriangleup\sigma)=0, \quad \dot\psi=0,\quad \ddot\sigma + \dot V - \phi + \psi =0, \quad 2\ddot\psi - \bigtriangleup\left(\ddot\sigma + \dot V - \phi + \psi\right) - \epsilon  \square (\phi-3\psi+\bigtriangleup\sigma)=0$$
which obviously reproduce our result (\ref{unimsym}) when $\epsilon\neq 0$.

At the same time, we do agree with the scalar-tensor representation \cite{unisym}, too. Indeed, one can either directly use the equations (\ref{feq3}) substituting every $Q_{\mu}$ coming with $\epsilon$ by $\partial_{\mu}\chi$ for the field $\chi\equiv \ln (-g)$, or (suppressing $\epsilon$) consider a Lagrangian system of
$${\mathcal L} =  {\mathbb Q} + g^{\mu\nu} (\partial_{\mu}\chi)(\partial_{\nu}\chi), \qquad \chi\equiv\ln (-g).$$
When we perform a variation of this action with respect to $g^{\mu\nu}$, in both its explicit appearance and through the field $\chi$, we get
$$ G_{\mu\nu} = T_{\mu\nu}(\chi=\ln (-g)) - g_{\mu\nu}\cdot \mathrm{e.o.m.}(\chi=\ln (-g)) = (\partial_{\mu}\chi)(\partial_{\nu}\chi) + g_{\mu\nu}\left(  2\mathop\square\limits^{(0)}\chi - \frac12(\partial\chi)^2\right)$$
with the superscript $(0)$ for Levi-Civita-related things. The Levi-Civita covariant divergence of the right hand side must be zero. In the covariant approach, it follows from the equation of motion for the connection. In our approach, it is anyway a self-consistency requirement \cite{meRM} since we work in vacuum, and the $G_{\mu\nu}$ part satisfies it.

The scalar field $\chi$ does not have its own equation of motion, for it was not an independent variable.  However, by finding the covariant divergence, we see
$$\left(\mathop\square\limits^{(0)}\chi\right) \partial_{\mu}\chi - 2\partial_{\mu}\mathop\square\limits^{(0)}\chi=0 $$
from which we get
$$\mathop\square\limits^{(0)}\chi = \lambda_0 e^{\chi/2}$$
with an arbitrary constant $\lambda_0$. Indeed, the field $\chi$ behaves precisely as it was stated in the Ref. \cite{unisym} and contributes the correct amount of effective energy-momentum to source the Einstein tensor.

What went wrong then? The count of degrees of freedom. If we take it as just the fully covariant GR with a dynamical scalar field possessing an arbitrary constant $\lambda_0$ in its Lagrangian, then it is indeed three dynamical modes, four constrained ones, and four pure gauges: the usual GR with a canonical scalar field. However, from the symmetric teleparallel perspective, there is also the flat connection, and the metric tensor is not a true tensor (unless we go for the covariant approach with more variables in it). The condition they have forgot to add is the requirement of
\begin{equation}
\label{unicond}
\chi=\ln (-g).
\end{equation}

In the logic of GR, it is just a choice of coordinates. However, we are not free in choosing the coordinates any longer. There are the preferred coordinates, the so-called coincident gauge ones. For a single GR solution, there might exist different flat structures on the manifold. As long as we take the flat connection as something objective and sensical, we must also take care of that. This is where the extra half degree of freedom comes from. Indeed, suppose we have found one solution, then there exist other solutions related to the initial one by a coordinate change $x^{\mu}\longrightarrow \zeta^{\mu}$ which, due to the condition (\ref{unicond}), must satisfy $\mathrm{det}\frac{\partial\zeta^{\mu}}{\partial x^{\nu}}=1$. Therefore, only three out of four former gauge freedoms are still pure gauge, while the fourth one becomes half-dynamical.

\subsection{On other modifications}

An interesting feature of the analysis above is that is does not show much difference in deviations from GR in terms of $a_3$ or $a_5$. In particular, we always see a half-integer number of degree of freedom, though the condition of non-degeneracy (\ref{heldeg}) is more intricate in terms of varying $a_5$ than that of varying $a_3$. It makes sense to try studying also the $a_5$-modified models, and not only the $a_3$-modified ones, for the latter can be constructed in line of unimodular gravity, and therefore do not require the symmetric teleparallel framework per se. At the same time, those two are the only parameters we might change in the linearised theory while keeping non-scalars potentially stable.

An interesting point to note is that, having taken a model of
$$0<a_1={\tilde a}_2=a_3, \qquad a_5=a_1 \cdot (1+\epsilon),$$
one can solve (\ref{hseq1}) for $\psi$ and substitute it in the kinetic quadratic form with the result of
$$\frac{2\epsilon(4+3\epsilon)}{(2+3\epsilon)^2}a_1 \left(\dot\phi+\bigtriangleup\dot\sigma\right)^2$$
which is non-negative for $\epsilon>0$. It does not really guarantee anything about stability by itself, though it looks very nice. We leave these models for future analysis.

\section{Conclusions}

We have analysed the models of Newer GR in the weak gravity limit. In this limit, they depend on only four free parameters out of five: $a_1$, ${\tilde a}_2 = a_2 - a_4$, $a_3$, and $a_5$. We cannot make the flat connection fully physical, in a sense that some parts of it must still be in pure gauge. This is because the first two parameters are immediately fixed to their GR values: even though $a_1> 0$ for presence of the usual tensors (gravitational waves), we then put ${\tilde a}_2 =a _1$, for otherwise there would be either ghosts or gradient instabilities in the vectors.  

Of course, the ghosts themselves do not produce any instability at the level of linearised theory (as long as the gradient energy has also got a reversed sign), and some may even (correctly) argue that ghosts might sometimes be all right \cite{vik1, vik}, see also recent developments \cite{Vero,vik2}. However, in case of gravity, it seems plausible that practically any situation which does not locally provide us with a reasonably stable ground state must lead to catastrophic outcomes. Nevertheless, more research of tamed ghosts should be done.

As we see it for now, there are not so many options of generalising STEGR with any kind of guaranteed stability. Some people find interest in using the GR-equivalent teleparallel models for defining the notions like conserved energy \cite{energ}. In our opinion, it is nothing but artificially endowing gravity with properties which do not naturally belong there. Therefore, it would be good to try as much of the new structure beyond the GR-equivalent options as possible.

For the scalars, after the restrictions from tensors and vectors, there remains freedom of choosing $a_3$ and $a_5$. The former is a safe option. It gives us the standard Einstein equation and a canonical scalar, with one half extra degree of freedom living in the flat connection which has not been noticed before \cite{unisym}. However, it does not make much of the symmetric teleparalell framework physical, for it can be constructed in a unimodular approach to gravity. In our opinion, the case of modifying $a_5$ is worth more study.

\end{document}